



\input lanlmac
\input amssym
\input epsf

\newcount\figno
\figno=0
\def\fig#1#2#3{
\par\begingroup\parindent=0pt\leftskip=1cm\rightskip=1cm\parindent=0pt
\baselineskip=13pt
\global\advance\figno by 1
\midinsert
\epsfxsize=#3
\centerline{\epsfbox{#2}}
\vskip 12pt
{\bf Fig. \the\figno:~~} #1 \par
\endinsert\endgroup\par
}
\def\figlabel#1{\xdef#1{\the\figno}}
\newdimen\tableauside\tableauside=1.0ex
\newdimen\tableaurule\tableaurule=0.4pt
\newdimen\tableaustep
\def\phantomhrule#1{\hbox{\vbox to0pt{\hrule height\tableaurule
width#1\vss}}}
\def\phantomvrule#1{\vbox{\hbox to0pt{\vrule width\tableaurule
height#1\hss}}}
\def\sqr{\vbox{%
  \phantomhrule\tableaustep

\hbox{\phantomvrule\tableaustep\kern\tableaustep\phantomvrule\tableaustep}%
  \hbox{\vbox{\phantomhrule\tableauside}\kern-\tableaurule}}}
\def\squares#1{\hbox{\count0=#1\noindent\loop\sqr
  \advance\count0 by-1 \ifnum\count0>0\repeat}}
\def\tableau#1{\vcenter{\offinterlineskip
  \tableaustep=\tableauside\advance\tableaustep by-\tableaurule
  \kern\normallineskip\hbox
    {\kern\normallineskip\vbox
      {\gettableau#1 0 }%
     \kern\normallineskip\kern\tableaurule}%
  \kern\normallineskip\kern\tableaurule}}
\def\gettableau#1 {\ifnum#1=0\let\next=\null\else
  \squares{#1}\let\next=\gettableau\fi\next}

\tableauside=1.0ex
\tableaurule=0.4pt


\def\Tr{{\rm Tr}}
\def\hf{{1\over 2}}

\def\o{\over}

\def\til#1{\widetilde{#1}}

\def\del{\partial}
\def\wg{\wedge}

\def\bra{\langle}
\def\ket{\rangle}
\def\lf{\left}
\def\ri{\right}
\def\riya{\rightarrow}

\def\lrya{\leftrightarrow}

\def\h#1{\widehat{#1}}

\def\dag{\dagger}
\def\rt#1{\sqrt{#1}}

\def\sitarel#1#2{\mathrel{\mathop{\kern0pt #1}\limits_{#2}}}

\lref\KashaniPoorXG{
  A.~K.~Kashani-Poor,
  ``Phase space polarization and the topological string: a case study,''
  arXiv:0812.0687 [hep-th].
}
\lref\DijkgraafFH{
  R.~Dijkgraaf, L.~Hollands and P.~Sulkowski,
  ``Quantum Curves and D-Modules,''
  arXiv:0810.4157 [hep-th].
}
\lref\Galetti{
D.~Galetti, 
``A Realization of the q-Deformed Harmonic Oscillator:
Rogers-Szeg\"{o} and Stieltjes-Wigert Polynomials,''
Braz. Jour. Phys. {\bf 33} 148 (2003).
}
\lref\KashaniPoorNC{
  A.~K.~Kashani-Poor,
  ``The Wave Function Behavior of the Open Topological String Partition
  Function on the Conifold,''
  JHEP {\bf 0704}, 004 (2007)
  [arXiv:hep-th/0606112].
}
\lref\OkuyamaEB{
  K.~Okuyama,
  ``D-brane amplitudes in topological string on conifold,''
  Phys.\ Lett.\  B {\bf 645}, 275 (2007)
  [arXiv:hep-th/0606048].
}
\lref\Eynard{
B. Eynard  and N. Orantin,
``Invariants of algebraic curves and topological expansion,''
[arXiv:math-ph/0702045].
}
\lref\HyunDR{
  S.~Hyun and S.~H.~Yi,
  ``Non-compact topological branes on conifold,''
  JHEP {\bf 0611}, 075 (2006)
  [arXiv:hep-th/0609037].
}
\lref\TierzJJ{
  M.~Tierz,
  ``Soft matrix models and Chern-Simons partition functions,''
  Mod.\ Phys.\ Lett.\  A {\bf 19}, 1365 (2004)
  [arXiv:hep-th/0212128].
}
\lref\OkudaMB{
  T.~Okuda,
  ``Derivation of Calabi-Yau crystals from Chern-Simons gauge theory,''
  JHEP {\bf 0503}, 047 (2005)
  [arXiv:hep-th/0409270].
}
\lref\IqbalDS{
  A.~Iqbal, N.~Nekrasov, A.~Okounkov and C.~Vafa,
  ``Quantum foam and topological strings,''
  arXiv:hep-th/0312022.
}
\lref\OkounkovSP{
  A.~Okounkov, N.~Reshetikhin and C.~Vafa,
  ``Quantum Calabi-Yau and classical crystals,''
  arXiv:hep-th/0309208.
}
\lref\SaulinaDA{
  N.~Saulina and C.~Vafa,
  ``D-branes as defects in the Calabi-Yau crystal,''
  arXiv:hep-th/0404246.
}
\lref\OoguriBV{
  H.~Ooguri and C.~Vafa,
  ``Knot invariants and topological strings,''
  Nucl.\ Phys.\ B {\bf 577}, 419 (2000)
  [arXiv:hep-th/9912123].
}
\lref\HalmagyiVK{
  N.~Halmagyi, A.~Sinkovics and P.~Sulkowski,
  ``Knot invariants and Calabi-Yau crystals,''
  JHEP {\bf 0601}, 040 (2006)
  [arXiv:hep-th/0506230].
}
\lref\GaiottoYB{
  D.~Gaiotto and L.~Rastelli,
  ``A paradigm of open/closed duality: Liouville D-branes and the  Kontsevich
  model,''
  JHEP {\bf 0507}, 053 (2005)
  [arXiv:hep-th/0312196].
}
\lref\MarinoFK{
  M.~Marino,
  ``Chern-Simons theory, matrix integrals, and perturbative three-manifold
  invariants,''
  Commun.\ Math.\ Phys.\  {\bf 253}, 25 (2004)
  [arXiv:hep-th/0207096].
}
\lref\AganagicWV{
  M.~Aganagic, A.~Klemm, M.~Marino and C.~Vafa,
  ``Matrix model as a mirror of Chern-Simons theory,''
  JHEP {\bf 0402}, 010 (2004)
  [arXiv:hep-th/0211098].
}
\lref\AganagicQJ{
  M.~Aganagic, R.~Dijkgraaf, A.~Klemm, M.~Marino and C.~Vafa,
  ``Topological strings and integrable hierarchies,''
  Commun.\ Math.\ Phys.\  {\bf 261}, 451 (2006)
  [arXiv:hep-th/0312085].
}
\lref\DijkgraafVP{
  R.~Dijkgraaf, A.~Sinkovics and M.~Temurhan,
  ``Universal correlators from geometry,''
  JHEP {\bf 0411}, 012 (2004)
  [arXiv:hep-th/0406247].
}
\lref\GopakumarVY{
  R.~Gopakumar and C.~Vafa,
  ``Topological gravity as large N topological gauge theory,''
  Adv.\ Theor.\ Math.\ Phys.\  {\bf 2}, 413 (1998)
  [arXiv:hep-th/9802016].
}
\lref\GopakumarKI{
  R.~Gopakumar and C.~Vafa,
  ``On the gauge theory/geometry correspondence,''
  Adv.\ Theor.\ Math.\ Phys.\  {\bf 3}, 1415 (1999)
  [arXiv:hep-th/9811131].
}
\lref\deHaroRZ{
  S.~de Haro and M.~Tierz,
  ``Discrete and oscillatory matrix models in Chern-Simons theory,''
  Nucl.\ Phys.\ B {\bf 731}, 225 (2005)
  [arXiv:hep-th/0501123].
}
\lref\Szego{
G. Szeg\"{o}, {\it Orthogonal Polynomials}, Colloquium Publications, Vol. 23
(American Mathematical Society).
}
\lref\FaddeevRS{
  L.~D.~Faddeev and R.~M.~Kashaev,
  ``Quantum Dilogarithm,''
  Mod.\ Phys.\ Lett.\ A {\bf 9}, 427 (1994)
  [arXiv:hep-th/9310070].
}
\lref\OoguriGX{
  H.~Ooguri and C.~Vafa,
  ``Worldsheet derivation of a large N duality,''
  Nucl.\ Phys.\ B {\bf 641}, 3 (2002)
  [arXiv:hep-th/0205297].
}
\lref\Walker{
P. J. Forrester, ``Vicious random walkers in the limit of a large number of walkers,'' J. Stat. Phys. {\bf 56}, 767 (1989).
}
\lref\WittenFB{
  E.~Witten,
  ``Chern-Simons gauge theory as a string theory,''
  Prog.\ Math.\  {\bf 133}, 637 (1995)
  [arXiv:hep-th/9207094].
}
\lref\WittenHF{
  E.~Witten,
  ``Quantum Field Theory And The Jones Polynomial,''
  Commun.\ Math.\ Phys.\  {\bf 121}, 351 (1989).
}
\lref\BershadskyCX{
  M.~Bershadsky, S.~Cecotti, H.~Ooguri and C.~Vafa,
  ``Kodaira-Spencer theory of gravity and exact results for quantum string
  amplitudes,''
  Commun.\ Math.\ Phys.\  {\bf 165}, 311 (1994)
  [arXiv:hep-th/9309140].
}
\lref\HoriKT{
  K.~Hori and C.~Vafa,
  ``Mirror symmetry,''
  arXiv:hep-th/0002222.
}
\lref\MarinoEQ{
  M.~Marino,
  ``Les Houches lectures on matrix models and topological strings,''
  arXiv:hep-th/0410165.
}
\lref\Andrews{
C. E. Andrews, {\it The Theory of Partitions} (Cambridge University Press,
1998).
}
\lref\AganagicGS{
  M.~Aganagic and C.~Vafa,
  ``Mirror symmetry, D-branes and counting holomorphic discs,''
  arXiv:hep-th/0012041.
}
\lref\AganagicNX{
  M.~Aganagic, A.~Klemm and C.~Vafa,
  ``Disk instantons, mirror symmetry and the duality web,''
  Z.\ Naturforsch.\ A {\bf 57}, 1 (2002)
  [arXiv:hep-th/0105045].
}
\lref\TierzVH{
  M.~Tierz,
  ``Chern-Simons theory, exactly solvable models and free fermions at finite
  temperature,''
  arXiv:0808.1079 [hep-th].
}
\lref\DolivetII{
  Y.~Dolivet and M.~Tierz,
  ``Chern-Simons matrix models and Stieltjes-Wigert polynomials,''
  J.\ Math.\ Phys.\  {\bf 48}, 023507 (2007)
  [arXiv:hep-th/0609167].
}
\lref\Atakishiyev{
N. M. Atakishiyev and Sh. M. Nagiyev,
``On the Rogers-Szego polynomials,''
J. Phys. A: Math. Gen. {\bf 27} L611 (1994).
}
\lref\DijkgraafSW{
  R.~Dijkgraaf, L.~Hollands, P.~Sulkowski and C.~Vafa,
  ``Supersymmetric Gauge Theories, Intersecting Branes and Free Fermions,''
  JHEP {\bf 0802}, 106 (2008)
  [arXiv:0709.4446 [hep-th]].
}
\lref\Szendroi{
B. Szendr\"{o}i,
``Non-commutative Donaldson-Thomas theory and the conifold,''
Geom. Topol. {\bf 12} 1171 (2008) [arXiv:0705.3419 [math.AG]].
}
\lref\VafaQF{
  C.~Vafa,
  ``Brane/anti-brane systems and $U(N|M)$ supergroup,''
  arXiv:hep-th/0101218.
}
\lref\OoguriYB{
  H.~Ooguri and M.~Yamazaki,
  ``Crystal Melting and Toric Calabi-Yau Manifolds,''
  arXiv:0811.2801 [hep-th].
}
\lref\ChuangAW{
  W.~y.~Chuang and D.~L.~Jafferis,
  ``Wall Crossing of BPS States on the Conifold from Seiberg Duality and
  Pyramid Partitions,''
  arXiv:0810.5072 [hep-th].
}
\lref\JafferisUF{
  D.~L.~Jafferis and G.~W.~Moore,
  ``Wall crossing in local Calabi Yau manifolds,''
  arXiv:0810.4909 [hep-th].
}
\lref\MaldacenaSN{
  J.~M.~Maldacena, G.~W.~Moore, N.~Seiberg and D.~Shih,
  ``Exact vs. semiclassical target space of the minimal string,''
  JHEP {\bf 0410}, 020 (2004)
  [arXiv:hep-th/0408039].
}
\lref\BouchardYS{
  V.~Bouchard, A.~Klemm, M.~Marino and S.~Pasquetti,
  ``Remodeling the B-model,''
  arXiv:0709.1453 [hep-th].
}

\Title{             
                                              }
{\vbox{
\centerline{$q$-Deformed Oscillators and D-branes on Conifold}
}}

\vskip .2in

\centerline{Kazumi Okuyama}
\vskip5mm
\centerline{Department of Physics, Shinshu University}
\centerline{Matsumoto 390-8621, Japan}
\centerline{\tt kazumi@azusa.shinshu-u.ac.jp}
\vskip .2in

\vskip 3cm
\noindent

We study the $q$-deformed oscillator algebra
acting on the wavefunctions
of non-compact D-branes in the topological string
on conifold. We find that the mirror B-model curve
of conifold appears from the commutation relation of
the $q$-deformed oscillators.

\Date{January 2009}

\vfill
\vfill
\newsec{Introduction}
The topological string
is an interesting playground to 
study the gauge/string duality via 
the geometric transition \refs{\GopakumarVY,\GopakumarKI,\OoguriGX}.
It is also interesting to study how the
target space geometry is quantized in
this context.
Recently, it is realized that the A-model side is
described by
a statistical model of crystal melting \refs{\OkounkovSP,\IqbalDS},
while 
the B-model side is reformulated 
as matrix models \refs{\Eynard,\BouchardYS}.
In both cases,
a spectral curve
appears either as the limit shape of molten crystal or
from the loop equation
of matrix model. 
It is expected that the spectral curve
should be viewed as a ``quantum Riemann surface''
in the sense that
the coordinates
of this curve become non-commutative at finite string
coupling $g_s$. It is argued that the natural language to deal with this
phenomenon is the $D$-module \refs{\DijkgraafSW,\DijkgraafFH}.

In this paper, we study the non-commutative algebraic structure
in the mirror B-model side of the topological string on the resolved
conifold 
${\cal O}(-1)\oplus{\cal O}(-1)\riya{\Bbb P}^1$.
As noticed in \DolivetII,
there is an underlying $q$-deformed oscillator 
(or, $q$-oscillator for short)
structure in the
wavefunction of non-compact D-branes on conifold. 
We study the representation of $q$-oscillators
in terms of non-commutative coordinates 
and show that the mirror curve of conifold
appears from the commutation relation of
the $q$-oscillators.

This paper is organized as follows. In section 2, we
construct the $q$-oscillators $A_{\pm}$
acting on the D-brane wavefunctions
in terms of variables obeying the commutation relation
$[p,x]=g_s$. In section 3, we show that the commutation relation
of $q$-oscillators $A_{\pm}$ is nothing but
the mirror curve of resolved conifold.
In section 4, we revisit the computation
of the partition function of
Chern-Simons theory using the $q$-oscillators.
We conclude in section 5 with discussion.

\subsec{Our Notations}
Here we summarize our notations and elementary formulas
used in the text.
We denote the string coupling as $g_s$,
and the K\"{a}hler parameter of the base ${\Bbb P}^1$
of resolved conifold
as $t=g_sN$. 
Then we introduce the parameters $q$ and $Q$ by
\eqn\defqQ{
q=e^{-g_s},\quad Q=q^N=e^{-t}.
}
We also introduce the canonical pair of coordinates $x$ and
$p$ satisfying
\eqn\pxcomm{
[p,x]=g_s.
} 
We use the representation such that $x$ acts as a multiplication
and $p$ acts as a derivative $p=g_s\del_x$.
In particular, when acting on a constant function ``1'',
we get
\eqn\xpactsonone{
x\cdot1=x,\quad p\cdot1=0,\quad e^{ax}\cdot1=e^{ax},\quad
e^{bp}\cdot1=1,
}
where $a$ and $b$ are c-number parameters. More generally,
$e^{bp}$ shifts $x$ when acting on a function $f(x)$
\eqn\ebponfx{
e^{bp}f(x)=f(x+bg_s).
}

We frequently use the commutation relations such as
\eqn\epxrel{
e^{ax+bp}=e^{ax}e^{bp}q^{-\hf ab},\quad
e^{bp}e^{ax}=e^{ax}e^{bp}q^{-ab},
}
which follow from the 
the relation $e^Ae^B=e^Be^Ae^{[A,B]}=e^{A+B}e^{\hf[A,B]}$
which is valid when $[A,B]$ is a c-number.

\newsec{Operator Representation of D-brane Wavefunctions}
In this section, we consider an operator representation
of the wavefunction of D-branes on conifold.
The wavefunction of a D-brane on conifold
in the standard framing is given by 
\refs{\OoguriBV\SaulinaDA\OkudaMB\HalmagyiVK\KashaniPoorNC\HyunDR{--}\KashaniPoorXG}
\eqn\ZNxBbrane{
Z_N(x)=\prod_{k=0}^\infty{1-q^{k+\hf}e^{-x}\o 1-Qq^{k+\hf}e^{-x}}
=\prod_{n=1}^N(1-q^{n-\hf}e^{-x})
=\sum_{r=0}^N\lf[\matrix{N\cr r}\ri]
q^{r^2\o2}(-1)^re^{-rx},
} 
where the $q$-binomial is defined as
\eqn\qbinom{
\lf[\matrix{N\cr r}\ri]={(q)_N\o(q)_{r}(q)_{N-r}},\quad
(q)_r=\prod_{n=1}^r(1-q^n).
}

In order to rewrite the wavefunction $Z_N(x)$ in the operator language,
let us recall the $q$-binomial formula 
for the variables $z$ and $w$ obeying the relation $wz=qzw$ 
\eqn\xyqbinomf{
(z+w)^N=\sum_{r=0}^N\lf[\matrix{N\cr r}\ri]z^rw^{N-r}.
}
Applying this formula for $z=-e^{-x+p}$ and $w=e^p$, we find
\eqn\Aplusbinomf{
(e^p-e^{-x+p})^N=\sum_{r=0}^N\lf[\matrix{N\cr r}\ri]
(-1)^re^{-rx+rp}e^{(N-r)p}
=\sum_{r=0}^N\lf[\matrix{N\cr r}\ri]
q^{r^2\o2}(-1)^re^{-rx}e^{Np}.
}
In the last step, we used the commutation relation \epxrel.
By comparing \ZNxBbrane\ and \Aplusbinomf, we see that
the D-brane wavefunction is written as
the operator $(e^p-e^{-x+p})^N$ acting on the constant
function ``1'', according to the rule in \xpactsonone.
Namely, the D-brane wavefunction has a simple expression
\eqn\ZNAplusN{
Z_N(x)=A_+^N\cdot1
}
where $A_+$ is given by
\eqn\Aplus{
A_+=e^p-e^{-x+p}.
}
Using the commutation relation \epxrel, $A_+$ is also written as
\eqn\Approdf{
A_+=(1-q^{\hf}e^{-x})e^p=e^p(1-q^{-\hf}e^{-x}).
}

One can see that the product form of $Z_N(x)$ in \ZNxBbrane\
easily follows from our simple expression $Z_N(x)=A_+^N\cdot1$.
By repeatedly using the relation \epxrel,
we can change the ordering so that
$e^p$ comes to the rightmost position
\eqn\mpowerAplus{\eqalign{
A_+^2&=(1-q^{\hf}e^{-x})e^p(1-q^{\hf}e^{-x})e^p=
(1-q^{\hf}e^{-x})(1-q^{\hf+1}e^{-x})e^{2p},\cr
A_+^3&=A_+^2
(1-q^{\hf}e^{-x})e^p=
(1-q^{\hf}e^{-x})(1-q^{\hf+1}e^{-x})
(1-q^{\hf+2}e^{-x})e^{3p},\cr
\cdots&\cdots\cr
A_+^N&=(1-q^{\hf}e^{-x})(1-q^{\hf+1}e^{-x})\cdots(1-q^{\hf+N-1}e^{-x})e^{Np}.
}}
When acting on ``1'', the last expression of $A_+^N$ gives
the product form of wavefunction in \ZNxBbrane.

\newsec{$q$-Oscillators and D-brane Wavefunctions}
In this section, we consider the $q$-oscillator
structure of the wavefunction $Z_N(x)$.
The $q$-oscillator structure
for the Rogers-Szeg\"{o} polynomials and the Stieltjes-Wigert
polynomials, which are related to our wavefunction $Z_N(x)$
by a change of framing \refs{\KashaniPoorNC,\HyunDR,\OkuyamaEB},
was studied in \refs{\Atakishiyev,\Galetti}.

From the definition $Z_N(x)=A_+^N\cdot 1$,
it follows that $A_+$ acts as the raising operator
\eqn\Znplusone{
A_+Z_N(x)=Z_{N+1}(x).
}  
Next consider the operator lowering the index of $Z_N(x)$.
From the relation 
\eqn\epZNact{
e^{-p}Z_N(x)=Z_{N}(x-g_s)=(1-q^{-\hf}e^{-x})Z_{N-1}(x),
}
and 
\eqn\ZNtoZNminusone{
Z_N(x)=(1-q^{N-\hf}e^{-x})Z_{N-1}(x),
}
one can see that
the operator $A_{-}$ defined by
\eqn\Aminusdef{
A_{-}={q^{\hf}e^{x}(1-e^{-p})\o 1-q},
}
lowers the index of $Z_N(x)$ as desired:
\eqn\AminusonZN{
A_{-}Z_N(x)={1-q^N\o1-q}Z_{N-1}(x).
}
Note that $A_-$ annihilates the constant function ``1''
\eqn\Amannone{
A_-\cdot1=0.
}
This implies that the constant function ``1'' can be identified as the
vacuum of $q$-oscillator
\eqn\oneasvac{
1~~\lrya~~|0\ket.
}

Now let us see that $A_{+}$ and $A_{-}$ obey the $q$-oscillator
algebra. From \Znplusone\ and \AminusonZN, we find
\eqn\ApmonZN{\eqalign{
A_+A_-Z_N(x)&={1-q^N\o1-q}A_+Z_{N-1}(x)={1-q^N\o1-q}Z_N(x), \cr
A_-A_+Z_N(x)&=A_-Z_{N+1}(x)={1-q^{N+1}\o1-q}Z_N(x).
}}
It follows that $A_+$ and $A_-$ satisfy the $q$-oscillator algebra
\eqn\Apmalg{
[A_-,A_+]=q^{\h{N}},\qquad A_-A_+-qA_+A_-=1.
}
Here the operator $q^{\h{N}}$ is defined as
\eqn\numberhN{
q^{\h{N}}A_{\pm}=q^{\pm1}A_{\pm}q^{\h{N}},\qquad q^{\h{N}}\cdot 1=1,
\qquad q^{\h{N}}Z_N(x)=q^NZ_N(x).
}

We can directly compute the algebra of $A_{\pm}$ using the
commutation relations \epxrel\
without acting them on the wavefunction as above.
From the  expression of $A_{\pm}$ in terms of variables $x,p$
in \Aplus\ and \Aminusdef, 
we can show that $A_{\pm}$ satisfy
\eqn\Apmprod{\eqalign{
A_-A_+&=1+{q(1-q^{-\hf}e^x)(1-e^p)\o1-q},\cr
A_+A_-&={(1-q^{-\hf}e^x)(1-e^p)\o1-q},\cr
[A_-,A_+]&=1-(1-q^{-\hf}e^x)(1-e^p),\qquad
A_-A_+-qA_+A_-=1.
}}
Therefore, we arrive at 
the expression of $q^{\h{N}}$ in terms of $x$ and $p$ as
\eqn\qhatNform{
q^{\h{N}}=1-(1-q^{-\hf}e^x)(1-e^p).
}
One can check that the RHS of \qhatNform\ satisfies
the defining properties of $q^{\h{N}}$ \numberhN.
When acting on $Z_N(x)$,
the relation \qhatNform\ leads to the following constraint on
the wavefunction $Z_N(x)$:
\eqn\constraintZN{
1-(1-q^{-\hf}e^x)(1-e^p)=Q.
}
This agrees with the known mirror B-model curve for the conifold 
\refs{\HoriKT,\AganagicWV}.
In other words, we find an interesting interpretation of
the mirror curve \constraintZN:
it represents the $q$-oscillator relation $[A_-,A_+]=q^{\h{N}}$
written in the canonical variables $x,p$.

\subsec{Wavefunction of anti-$D$-Brane}
In contrast to the ordinary oscillator,
in the case of $q$-oscillator 
we can consider the formal inverse of $A_+=e^p(1-q^{-\hf} e^{-x})$
\eqn\Apinv{
A_+^{-1}=(1-q^{-\hf}e^{-x})^{-1}e^{-p}.
}
Repeating the similar calculation as in \mpowerAplus, we find
\eqn\ApminusN{\eqalign{
A_+^{-N}&=(1-q^{-\hf}e^{-x})^{-1}e^{-p}(1-q^{-\hf}e^{-x})^{-1}\cdots
(1-q^{-\hf}e^{-x})^{-1}e^{-p} \cr
&=(1-q^{-\hf}e^{-x})^{-1}(1-q^{-\hf-1}e^{-x})^{-1}\cdots
(1-q^{-\hf-N+1}e^{-x})^{-1}e^{-Np}\cr
&=\prod_{n=1}^N(1-q^{n-\hf-N}e^{-x})^{-1}e^{-Np}.
}}
From this expression, we see that  
\eqn\DbarZN{
Z_{-N}(x)=A_+^{-N}\cdot1={1\o Z_N(x-t)}.
}
As argued in \SaulinaDA,
this is interpreted as the wavefunction of
anti-D-brane, up to a shift of $x$. $Z_{-N}(x)$ has another interpretation 
as the wavefunction of D-brane ending on a different leg of the toric
diagram of conifold \refs{\KashaniPoorNC,\HyunDR,\KashaniPoorXG}. 
The constraint equation $[A_-,A_+]=q^{\h{N}}$ for $Z_{-N}(x)$ reads
\eqn\tZNwaveeq{
1-(1-q^{-\hf}e^{x})(1-e^{p})=Q^{-1},
}
which can be rewritten as
\eqn\eqfortldex{
1-(1-q^{-\hf}e^{x-t})(1-e^{-p})=Q.
}
This is the same form as the equation for $Z_N(x)$ \constraintZN\
under the replacement
$(x,p)\riya (x-t,-p)$.
This is consistent with the wavefunction behavior of
D-brane amplitude under the change of polarization
\refs{\AganagicQJ,\KashaniPoorNC}.

\newsec{Closed String Partition Function and the $q$-Oscillators}
In this section, we consider the  partition
function of closed topological string on conifold
from the viewpoint of $q$-oscillators. Although our computation
is essentially the same as \TierzJJ, we emphasize that
the $q$-oscillator structure makes the computation
more transparent.
There is essentially
no new result in this section, but we include this
for completeness. 

As shown in \refs{\WittenFB,\GopakumarKI,\GopakumarVY},
the closed string 
partition function of conifold is given by
the $U(N)$ Chern-Simons theory on $S^3$.
Later, it was noticed in \TierzJJ\
that the same partition function
is written as the log-normal matrix model
\eqn\Zlogdef{
Z_{\rm log}=\int_{N\times N} dM e^{-{1\o2g_s}\Tr(\log M)^2}.
}
The orthogonal polynomial associated with this log-normal measure
is known as the Stieltjes-Wigert polynomial $S_N(x)$ \Szego,
which is given by
\eqn\SNxform{
S_N(x)=\sum_{k=0}^N\lf[\matrix{N\cr k}\ri]q^{k^2+\hf k}(-1)^ke^{-kx}.
}

In the following we will show that $S_N(x)$ is related to 
$Z_N(x)$ by the following
change of framing
\eqn\frameSW{
x\riya x-p+\hf g_s,\quad p\riya p.
}
This is realized by the conjugation 
by the operator $U=e^{-{p^2\o2g_2}+\hf p}$
\eqn\conjbyepsq{
UxU^{-1}=x-p+\hf g_s,\quad
UpU^{-1}=p.
}
Note that the operator $U$ preserves the vacuum
\eqn\Uonvac{
U\cdot1=1~~\lrya~~U|0\ket=|0\ket.
}
In terms of the conjugated $q$-oscillators
\eqn\Bpmdef{\eqalign{
B_+&=UA_+U^{-1}=e^p-q^{\hf}e^{-x+2p}=e^p-q^{3\o2}e^{-x}e^{2p}\cr
B_-&=UA_-U^{-1}={q^{\hf}e^x(e^{-p}-e^{-2p})\o 1-q}
}}
the Stieltjes-Wigert polynomial is written in
the same form as $Z_N(x)=A_+^N\cdot1$
\eqn\SNinBpN{
S_N(x)=B_+^N\cdot1.
}
Using the commutation relation \epxrel,
one can see that \SNinBpN\ agrees with the expression \SNxform,
as promised
\eqn\SWinBp{
S_N(x)=\sum_{k=0}^N\lf[\matrix{N\cr k}\ri]
(-q^{\hf}e^{-x+2p})^ke^{(N-k)p}\cdot1 
=\sum_{k=0}^N\lf[\matrix{N\cr k}\ri]q^{k^2+\hf k}(-1)^ke^{-kx}.
}

In order to calculate the partition function \Zlogdef,
we need the norm of the function $\Psi_N(x)$ defined by
\eqn\FZZT{
\Psi_N(x)\equiv\big\bra \det(e^{-x}-M)\big\ket=(-1)^Nq^{-N^2-{N\o2}}S_N(x),
}
which is known as the FZZT wavefunction \refs{\AganagicQJ,\MaldacenaSN}.
We see that the $q$-oscillator representation
\SNinBpN\ simplify the computation
of the norm.

The log-normal measure associated with
\Zlogdef\ becomes Gaussian under the change of variable $y=e^{-x}$
\eqn\mesdmu{
\int_0^\infty {dy\o2\pi}e^{-{1\o2g_s}(\log y)^2}
=\int_{-\infty}^\infty{dx\o2\pi}e^{-{x^2\o2g_s}-x}.
}
The inner product with respect to this measure is defined as
\eqn\innerfg{
\bra f,g\ket\equiv\int_{-\infty}^\infty{dx\o2\pi}e^{-{x^2\o2g_s}-x}
f(x)g(x).
}
Let us consider the adjoint of $B_+$ with respect to this measure
\eqn\Bdagdefinn{
\bra f,B_+g\ket=\bra (B_+)^\dag f,g\ket.
}
Using the representation of $B_+$ in terms of
$x,p$ \Bpmdef, the action of $B_+$ on a function $g(x)$ reads
\eqn\Bponfofx{
B_+g(x)=g(x+g_s)-q^{3\o2}e^{-x}g(x+2g_s).
}
Then the inner product $\bra f,B_+g\ket$ is written as
\eqn\Bpadcomp{\eqalign{
\bra f,B_+g\ket&=\int_{-\infty}^\infty{dx\o2\pi}e^{-{x^2\o2g_s}-x}
f(x)\Big[g(x+g_s)-q^{3\o2}e^{-x}g(x+2g_s)\Big] \cr
&=\int_{-\infty}^\infty{dx\o2\pi}e^{-{x^2\o2g_s}-x}
q^{-\hf}e^x\Big[f(x-g_s)-f(x-2g_s)\Big]g(x). 
}}
From this equation, we find that the adjoint of $B_+$ is
proportional to $B_-$ \Bpmdef
\eqn\adjBptoBm{
(B_+)^\dag=q^{-\hf}e^{x}(e^{-p}-e^{-2p})=(q^{-1}-1)B_-.
}

Now it is straightforward to calculate the norm
of $S_N(x)$. In order to do that, it is convenient
to use the bra-ket notation
\eqn\SnxtoQM{\eqalign{
1~~&\lrya~~|0\ket \cr
S_n(x)=B_+^n\cdot1~~&\lrya~~|n\ket=B_+^n|0\ket.
}}
Noticing that
$B_-$ satisfies the same relation as $A_-$ \AminusonZN\
when acting on the state $|n\ket$
\eqn\Bmonnket{
B_-|0\ket=0,\quad B_-|n\ket={1-q^n\o1-q}|n-1\ket,
}
and using the 
relation between $(B_+)^\dag$ and $B_-$ \adjBptoBm,
we find
\eqn\Badjonnket{
(B_+)^\dag|n\ket=q^{-1}(1-q^n)|n-1\ket.
}

Now let us compute the norm $\bra n|n\ket$.
First, the norm of unit function $1=|0\ket$ is given by
\eqn\zeronorm{
\bra0|0\ket=\int_{-\infty}^\infty{dx\o2\pi}e^{-{x^2\o2g_s}-x}
=\rt{g_s\o2\pi}q^{-\hf}.
}
The norm of higher state $|n\ket$ is determined by the following
recursion relation
\eqn\normnketrecursion{
\bra n|n\ket=\bra n-1|(B_+)^\dag|n\ket=
q^{-1}(1-q^n)\bra n-1|n-1\ket.
}
Finally, the norm of $|n\ket$ is found to be
\eqn\normnketprod{
\bra n|n\ket=\bra0|0\ket q^{-n}\prod_{k=1}^n(1-q^k)
=\rt{g_s\o2\pi}q^{-n-\hf}\prod_{k=1}^n(1-q^k).
}
This agrees with the
known result of the norm of $S_n(x)$ with respect to the measure
\mesdmu\ \Szego.

The partition function of matrix model
\Zlogdef\ is given by the product of the
norm of $\Psi_n(x)$ in \FZZT
\eqn\hnnorm{
h_n=\int_{-\infty}^\infty{dx\o2\pi}e^{-{x^2\o2g_s}-x}\Psi_n(x)^2=
q^{-2n^2-n}\bra n|n\ket=\rt{g_s\o2\pi}q^{-2(n+\hf)^2}\prod_{k=1}^n(1-q^k).
}
Then the 
partition function of matrix model
\Zlogdef\ is given by
\eqn\Zlogprod{
Z_{\rm log}
=\prod_{n=0}^{N-1}h_n
=\eta(\til{q})^Nq^{-{2N^3\o3}+{N\o8}}\prod_{n=1}^\infty
\lf({1-Qq^n\o1-q^n}\ri)^n
}
where $\eta(\til{q})$ denotes the $\eta$-function
\eqn\etatilde{
\eta(\til{q})=\rt{g_s\o2\pi}q^{1\o24}\prod_{n=1}^\infty(1-q^n)
=\til{q}^{1\o24}\prod_{n=1}^\infty(1-\til{q}^n),
}
with
\eqn\qtilde{
\til{q}=e^{-{4\pi^2\o g_s}}.
}

On the other hand, the partition function
of the $U(N)$ Chern-Simons theory on $S^3$ is \WittenHF
\eqn\ZCSpart{
Z_{CS}=\lf({g_s\o2\pi}\ri)^{N\o2}\prod_{k=1}^{N-1}(q^{-{k\o2}}-q^{k\o2})^{N-k}
=\eta(\til{q})^Nq^{-{N^3\o12}+{N\o24}}\prod_{n=1}^\infty
\lf({1-Qq^n\o1-q^n}\ri)^n,
}
which agrees with $Z_{\rm log}$ up to terms in the free energy which
are polynomial in $t$.

We note in passing that $Z_{\rm log}$ is written as the norm
of Fermi sea state $|\Psi\ket$
\eqn\ZloginPsi{
Z_{\rm log}\propto\bra\Psi|\Psi\ket~,\qquad
|\Psi\ket={1\o\rt{N!}}|0\ket\wg|1\ket\wg\cdots\wg|N-1\ket.
}

\newsec{Discussion}
In this paper, we find that the mirror B-model curve of
resolved conifold has an interesting interpretation
as the $q$-oscillator relation $[A_-,A_+]=q^{\h{N}}$ itself.
It would be interesting to find the physical origin
of this algebraic structure.

Recently, the partition function of 
the Donaldson-Thomas theory
of the non-commutative version of
conifold is calculated by Szendr\"{o}i \Szendroi
\eqn\Znoncom{
Z_{NC}=\prod_{n=1}^\infty\lf({1-Q^{-1}q^n\o1-q^n}\ri)^n
\prod_{n=1}^\infty\lf({1-Qq^n\o1-q^n}\ri)^n.
}
The last factor of
$Z_{NC}$ in \Znoncom\
is the same as the Chern-Simons partition function
in \ZCSpart, but the first factor is different
\eqn\etaandQ{
\eta(q)^N=q^{N\o24}\prod_{k=1}^\infty(1-q^k)^N=
q^{N\o24}\prod_{n=N+1}^\infty(1-Q^{-1}q^n)^n
\prod_{n=1}^\infty(1-q^n)^{-n}.
}
This difference is discussed 
in the context of the wall-crossing phenomena 
\refs{\JafferisUF,\ChuangAW,\OoguriYB}.
It is tempting to identify the
extra factor in $Z_{NC}$ as the effect of anti-D-branes \VafaQF
\eqn\missingF{
\prod_{n=1}^N(1-Q^{-1}q^n)^n\sim\prod_{n=1}^N{1\o\bra -n|-n\ket}.
}
However, one should not take this relation literally, since
both sides of \missingF\ vanish when $N$ is an integer.
We leave this as an interesting future problem.

\vskip10mm
\noindent
\centerline{\bf Acknowledgments}
I would like to thank  A-K. Kashani-Poor, S. Odake, and 
Ta-Sheng Tai for discussion and correspondence. 
This work is supported in part by
MEXT Grant-in-Aid for Scientific Research \#19740135.
\listrefs
\bye